\documentstyle[12pt,a4]{article}

\newcommand{\VEV}[1]{\left\langle #1 \right\rangle}

\newcommand{\reseteqnum}{\setcounter{equation}{0}}
\newcommand{\rbox}[1]{\vbox{\hrule height.8pt%
                \hbox{\vrule width.8pt\kern5pt
                \vbox{\kern5pt\hbox{#1}\kern5pt}\kern5pt
                \vrule width.8pt}
                \hrule height.8pt}}
\begin{document}
\renewcommand{\thepage}{}

\begin{titlepage}
\title{
\hfill
\parbox{4cm}{\normalsize KUNS-1372 \\HE(TH)~95/22\\
hep-th/9512015}\\
\vspace{4ex}
Conformal Anomaly of String Theory\vspace{2ex}\\
in the Harmonic Gauge
\vspace{9ex}}

\author{Tomohiko Takahashi\thanks{e-mail address:
 {\tt tomo@gauge.scphys.kyoto-u.ac.jp}}
\thanks{JSPS Research Fellow}
\vspace{3ex}\\
  {\it Department of Physics, Kyoto University}\\
   {\it Kyoto 606-01, Japan}}
\date{December, 1995}

\maketitle
\vspace{8ex}

\begin{abstract}
\normalsize
\baselineskip=19pt plus 0.2pt minus 0.1pt
Considering the conformal anomaly in an effective action, the
critical dimension of string theory can be decided in the harmonic
gauge, in which it had been reported before to be indefinite. In this
gauge, there is no anomaly for the ghost number symmetry. This can be
naturally understood in terms of Faddeev-Popov conjugation in the theory.
\end{abstract}
\end{titlepage}

\renewcommand{\thepage}{\arabic{page}}
\setcounter{page}{1}
\baselineskip=19pt plus 0.2pt minus 0.1pt
\section{Introduction}
\reseteqnum
\input epsf
\noindent
One of the interesting features of string theories is the existence of
the critical dimension, at which a lot of fascinating
events occur, including: the appearance of massless spin-one and spin-two
states, the
realization of Lorentz covariance in the light-cone gauge\,\cite{rf:GGRT}, the
nilpotency
of the BRS charge in the conformal gauge\,\cite{rf:KatOga}, and the
cancellation
of the conformal
anomalies\,\cite{rf:Poly}.

Most of these subjects have been investigated in the conformal or
light-cone gauge, whereas several works have been done in the harmonic
gauge. The
first-quantized bosonic strings in the harmonic gauge have a relation to
$OSp(1,1|\,2)$ string field theory\,\cite{rf:BSZ,rf:SieZwi}.
In the latter theory, which is a free one,
the $OSp(1,1|\,2)$ algebra
with the BRS and anti-BRS generators has been shown to close at $D=26$ only.
Besides, the cancellation of the conformal
anomaly has been studied in the harmonic gauge
from a Lagrangian approach, and this occurs at the same
dimension\,\cite{rf:Duse}. There are also other arguments in different
versions of the harmonic gauge\,\cite{rf:FRP,rf:KR}.

In general, it is believed that the critical dimension is a well-defined
concept. However, there have been a few
contradictory results, {\it e.g.},
indefiniteness of the critical
dimension in the harmonic gauge\,\cite{rf:AbeNaka}. In this gauge, the
dependence of a gauge parameter appears in the two-point function
of the energy momentum tensor based on a perturbative calculation. The
anomalous term of the two-point function would have been
expected to vanish at $D=26$, as in any other
gauge, but it does not, due to its gauge dependence.
This result suggests that the anomaly may disappear for any $D$ by
adjusting the gauge parameter.
There is also a study in the temporal gauge indicating that the critical
dimension may not be determined as far as considering a cylinder
amplitude\,\cite{rf:Kawano}.

In the following, we will decide the critical dimension in the harmonic
gauge. In order to clarify the above discrepancy, it is necessary
to analyze the anomaly more carefully.
We will consider an effective action and the Ward-Takahashi identity
for the decision of the critical dimension.

Moreover, we
will study the ghost number anomaly. In the conformal gauge, the ghost
number current has the anomaly\,\cite{rf:fuji}, the integrated version
of which can be
interpreted as an index theorem. Contrastingly, it has been known
that the ghost number anomaly does not exist in the harmonic
gauge\,\cite{rf:Duse}. We will compute
perturbatively the ghost
number anomaly in the harmonic gauge with a gauge parameter considered
throughout this paper.

\section{Harmonic Gauge}
\reseteqnum

\noindent
The Lagrangian of two dimensional gravity coupled to matters in
the harmonic gauge\,\cite{rf:NakaOji} has the form,
\begin{eqnarray}
{\cal L} = -\frac{1}{2}\widetilde{g}^{\alpha \beta}
      \partial_\alpha X_M \partial_\beta X^M
      -\widetilde{g}^{\alpha \beta} \partial_\alpha b_\beta
    - i \partial_\alpha \overline{c}_\beta \left\{
      \widetilde{g}^{\alpha \gamma}\partial_\gamma c^\beta
      +\widetilde{g}^{\beta \gamma}\partial_\gamma c^\alpha
      -\left(\partial_\gamma \widetilde{g}^{\alpha \beta}\right)
      c^\gamma \right\},
\end{eqnarray}
where $\widetilde{g}^{\alpha \beta} = \sqrt{-g} g^{\alpha \beta}$ and
Greek indices correspond to world sheet coordinates, and
roman indices
run from 1 to $D$. $b_\alpha$, $c^\alpha$ and $\overline{c}_\alpha$
denote the auxiliary field, the ghost and anti-ghost fields,
respectively. It
is possible to obtain a more simplified Lagrangian
by the redefinition of the field variables; the shift of the auxiliary
field $b_\alpha$\,\cite{rf:Naka},
\begin{eqnarray}
b_\alpha = \widetilde{b}_\alpha + i\,(1-\xi)\,c^\beta \partial_\beta
           \overline{c}_\alpha.
\end{eqnarray}
The Jacobian for this transformation is trivial and so
there is no anomaly in it. After performing this shift and
dropping a total
derivative term, we find a Lagrangian as follows,
\begin{eqnarray}
{\cal L} = -\frac{1}{2}\widetilde{g}^{\alpha \beta}
      \partial_\alpha X_M \partial_\beta X^M
      -\widetilde{g}^{\alpha \beta}
        \partial_\alpha \widetilde{b}_\beta
      - i \widetilde{g}^{\alpha \beta}
     \partial_\alpha \overline{c}_\gamma \partial_\beta c^\gamma
   +i\,\xi\,\widetilde{g}^{\alpha \beta}\,\partial_\alpha
      \left(\,c^\gamma\,\partial_\gamma\overline{c}_\beta\right),
    \label{eq:lag}
\end{eqnarray}
where $\xi$ is a gauge parameter.
In the case of $\xi=0$, the Lagrangian has a simpler form than
before. In this paper, we will quantize the theory defined by the latter
Lagrangian and discuss anomalies derived from it.

The action is invariant under the BRS symmetry:
\begin{eqnarray}
&&\delta_B X^M  =  -c^\alpha \partial_\alpha X^M,
\hspace{2.5em}
\delta_B g^{\alpha \beta} = g^{\alpha \gamma}\partial_\gamma c^\beta
+ g^{\beta \gamma}\partial_\gamma c^\alpha -
\left(\partial_\gamma g^{\alpha \beta}\right)c^\gamma,
\nonumber\\
&&\delta_B c^\alpha  =  -c^\beta \partial_\beta c^\alpha,
\hspace{4em}
\delta_B \overline{c}_\alpha  =  i \widetilde{b}_\alpha
         -(1-\xi)\,c^\beta \partial_\beta \overline{c}_\alpha,
\nonumber\\
&&\delta_B \widetilde{b}_\alpha  =
   -(1-\xi)\,c^\beta \partial_\beta \widetilde{b}_\alpha.
\label{eq:brstrans}
\end{eqnarray}
Moreover the Faddeev-Popov(FP) conjugate transformation also leaves it
invariant:
\begin{eqnarray}
{\cal C}_{FP}\ :\ \ c^\alpha &\longrightarrow& \eta^{\alpha \beta}
              \overline{c}_\beta, \nonumber\\
 \overline{c}_\alpha &\longrightarrow& -\eta_{\alpha \beta}c^\beta,
\nonumber\\
\widetilde{b}_\alpha &\longrightarrow& \widetilde{b}_\alpha +
i\,\xi\,\left(\,
\partial_\gamma \overline{c}_\alpha\,c^\gamma +
\eta_{\alpha \beta}\,\eta^{\gamma \delta}\,\partial_\gamma c^\beta
\,\overline{c}_\delta\,\right),
\label{eq:fpconj}
\end{eqnarray}
where $\eta^{\alpha \beta}$ is the world sheet flat metric. Therefore
we find out the extended BRS symmetry, which involves the
anti-BRS and the ghost number symmetry in addition to the BRS
symmetry.

Now let us consider the case of parameterizing the metric as follows,
\begin{eqnarray}
g^{\alpha \beta} = e^\phi \left(\eta^{\alpha \beta}
                       - h^{\alpha\beta}\right), \label{eq:exp}
\hspace{2em}\eta_{\alpha \beta} h^{\alpha \beta} = 0, \label{eq:expmet}
\end{eqnarray}
where $\eta^{\alpha \beta}={\rm diag}(+1,-1)$. The degrees of freedom
of $g^{\alpha
\beta}$ are the
same as the ones of $h^{\alpha \beta}$ and $\phi$:
because of the tracelessness of $h^{\alpha \beta}$, it has two degrees of
freedom, and $\phi$ has one. By substituting the expression of
Eq.~(\ref{eq:exp}) to Eq.~(\ref{eq:lag}), we obtain the following Lagrangian,
\begin{eqnarray}
&&{\cal L} \equiv{\cal L}_0 + {\cal L}_{int}, \nonumber\\\vspace{10ex}
&&{\cal L}_0 \equiv -\frac{1}{2}\,\partial_\alpha X_M\,\partial^\alpha
X^M
   + h^{\alpha \beta}\,\partial_\alpha \widetilde{b}_\beta
   - i\,\partial_\alpha \overline{c}_\beta\,\partial^\alpha c^\beta,
\nonumber\\
&&{\cal L}_{int} \equiv \frac{1}{2}\,h^{\alpha \beta}\,\partial_\alpha
X_M\,
                     \partial_\beta X^M
                  -\frac{1}{4}\,h_{\alpha \beta}\,h^{\alpha \beta}\,
                   \partial^\gamma \widetilde{b}_\gamma
             + i\, h^{\alpha \beta}\,\partial_\alpha \overline{c}_\gamma\,
                  \partial_\beta c^\gamma
             +i\,\xi\,h^{\alpha \beta}\,\partial_\alpha
        \left(\,\partial_\gamma \overline{c}^{}_\beta\,c^\gamma\,
        \right)
      \nonumber\\
   && \hspace{3em}-\frac{1}{8}\,h_{\alpha \beta}\,h^{\alpha \beta}
        \partial_\gamma X_M\, \partial^\gamma X^M
       + \frac{1}{4}\, h_{\alpha \beta}\, h^{\alpha \beta}\,
         h^{\gamma \delta}\, \partial_\gamma \widetilde{b}_\delta
       - \frac{i}{4}\,h_{\alpha \beta}\,h^{\alpha \beta}\,
         \partial_\gamma \overline{c}_\delta\,\partial^\gamma c^\delta
\nonumber\\
 && \hspace{3em} -\frac{i}{4}\,\xi\,h^{\alpha \beta}\,h_{\alpha \beta}\,
      \partial_\gamma
\left(\,\partial_\delta \overline{c}^\gamma\,c^\delta\,\right)
       + \cdots,
\end{eqnarray}
where the tensor indices are raised or lowered by the flat metric
$\eta^{\alpha \beta}$, and the ellipsis denotes terms of higher order
than the fourth power of the fields. As a consequence of the conformal
symmetry
remaining in the Lagrangian of Eq.~(\ref{eq:lag}), there is no
dependence of the conformal mode $\phi$ in it. This is the same
situation as in the conformal or light-cone gauge. Though the gauge fixing
has not yet been
performed for the conformal symmetry, it is not needed for our
discussion.

Propagators are derived from the free part of the Lagrangian:
\begin{eqnarray}
&&\VEV{{\rm T~} X^M(x)\,X^N(y)}_0 = g^{M N} \int \frac{d^2p}{i (2 \pi)^2}\,
  \frac{1}{p^2 + i \epsilon}\,{\rm e}^{-ipx},  \nonumber\\
&&\VEV{{\rm T~} c^\alpha(x)\,\overline{c}_\beta(y)}_0 =
 \delta^\alpha_\beta \int \frac{d^2p}{i (2
\pi)^2}\,\frac{-i}{p^2+i\epsilon}\,
 {\rm e}^{-ipx},  \nonumber\\
&&\VEV{{\rm T~} \widetilde{b}_\alpha(x)\,h_{\beta \gamma}(y)}_0 =
 \int \frac{d^2p }{i (2 \pi)^2}\,
 \frac{-i}{p^2+i\epsilon}\,\left(p_\beta \eta_{\alpha \gamma}
 + p_\gamma \eta_{\alpha \beta}-p_\alpha \eta_{\beta \gamma}\right)\,
  {\rm e}^{-ipx},
\end{eqnarray}
where we take $\VEV{\ }_0$ to correspond to a free propagator.

\begin{figure}[t]
\unitlength 2.58em
\begin{picture}(13,6)(0,-0.3)
%
\unitlength 2.58em
\put(4.3,5.5){$=\VEV{X^M\,X^N}_0$}
\put(4.3,4.75){$=\VEV{c^\alpha\,\overline{c}_\beta}_0$}
\put(4.3,3.95){$=\VEV{\widetilde{b}_\alpha\,h_{\beta \gamma}}_0$}
\put(0.05,2.95){$X_M(p)$}
\put(0.05,1.9){$X_N(q)$}
\put(2.5,2.8){$h_{++}$}
\put(4.2,2.3){$-{\displaystyle \frac{1}{2}}\,g^{MN}\,p_-q_-$}
\put(0.2,1.1){$\overline{c}_+(p)$}
\put(0.2,0){$c_-(q)$}
\put(2.5,0.9){$h_{++}$}
\put(4.2,0.4){$-\,i\,p_-q_-$}
\put(7.2,5.5){$\overline{c}_-(p)$}
\put(7.2,4.35){$c_+(p)$}
\put(9.5,5.3){$h_{++}$}
\put(9.3,4.4){$-\,i\,p_-q_- -\,i\,\xi\,(p_-+q_-)p_-$}
\put(7.2,3.4){$\overline{c}_-(p)$}
\put(7.2,2.3){$c_-(q)$}
\put(9.5,3.2){$h_{++}$}
\put(10,2.35){$-\,i\,\xi\,(p_-+q_-)p_-$}
\put(7.2,1.3){$\widetilde{b}_+(p)$}
\put(7.2,0){$h_{--}$}
\put(9.7,0.9){$h_{++}$}
\put(11.5,0.4){${\displaystyle \frac{i}{2}}\,p_-$}
\hspace{3em}
\epsfxsize=25em
\epsfbox{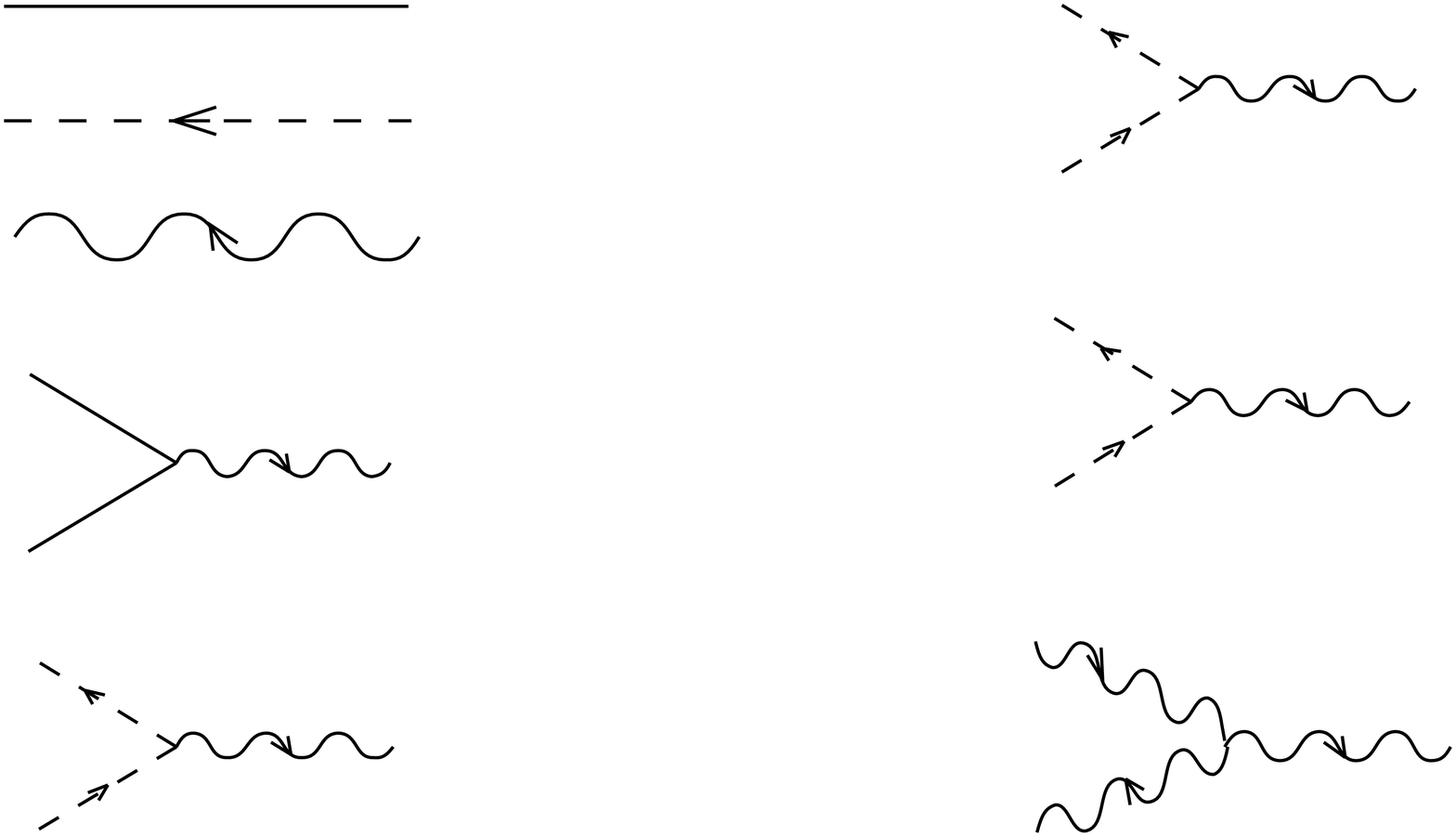}
\end{picture}
\caption{Graphical representation of propagators and vertices. Higher
order vertices are omitted. The vertices with $+$ and $-$
components exchanged also exist.}
\label{fig:Fey}
\end{figure}

It is convenient, for practical calculation, to introduce light-cone
coordinates,
\begin{eqnarray}
x^{\pm} = \frac{1}{\sqrt{2}} \left(x^0 \pm x^1\right).
\end{eqnarray}
The associated metric is defined as
$\eta^{+-}=\eta^{-+}=\eta_{+-}=\eta_{-+}=1$ and otherwise zero.
In Fig.~\ref{fig:Fey} we have represented the resultant Feynman rule,
which is expressed with these coordinates.

Using this perturbative method, we will compute the
two-point function of the energy momentum tensor.
It is given by
\begin{eqnarray}
T_{\alpha \beta} &\equiv& -\frac{2}{\sqrt{-g}}\frac{\delta S}{\delta
g^{\alpha \beta}}
\nonumber\\
&=&\partial_\alpha X_M\,\partial_\beta X^M
 + \partial_{( \alpha}\,\widetilde{b}_{\beta)}
 + i\,\partial_{(\alpha}
\overline{c}_\gamma\,\partial_{\beta)}c^\gamma
 -i\,\xi\,\partial_{(\alpha}\left(c^\gamma\,\partial_\gamma
   \overline{c}_{\beta)}\right)
 \nonumber\\
&& - g_{\alpha \beta}\,g^{\gamma
\delta}\,\left(\,\frac{1}{2}\,\partial_\gamma
   X_M\,\partial_\delta X^M + \partial_\gamma\,\widetilde{b}_\delta
 +i\,\partial_\gamma \overline{c}_\eta\,\partial_\delta c^\eta
  -i\,\partial_\gamma\left(c^\eta\,\partial_\eta\overline{c}_\delta\,\right)
 \right).
\end{eqnarray}
It satisfies
\begin{eqnarray}
T_{\alpha \beta} = 0.  \label{eq:eqmotion}
\end{eqnarray}
In the harmonic gauge, Eq.~(\ref{eq:eqmotion}) gives the
equation of motion for $\widetilde{b}_\alpha$ perturbatively, unlike in
the conformal or the light-cone gauge, in which it corresponds to the Virasoro
constraint.

\begin{figure}[t]
\unitlength 2.8em
\begin{picture}(13,1.6)
\put(4.95,0.6){$T_{++}(p)$}
\put(7.95,0.6){$T_{++}(-p)$}
\centerline{
\epsfxsize=4.3em
\epsfbox{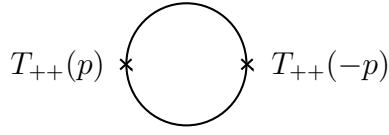}
}
\end{picture}
\caption{A Feynman diagram evaluated in Eq.~(2.12) as an example
calculation.}
\label{fig:enerX}
\end{figure}

We illustrate the evaluation of the Feynman diagram in
Fig.~\ref{fig:enerX}\,\,\cite{rf:AlWitt,rf:tHooft}. It is given by
\begin{eqnarray}
&&\hspace{2em}-\frac{D}{2} \int \frac{dk_+ dk_-}{i (2 \pi)^2}\,
   \frac{k_+ k_+}{k_+ k_- +i \epsilon}\,
   \frac{(k_+ -p_+)(k_+ -p_+)}{(k_+ -p_+)(k_- -p_-)+i \epsilon}
\nonumber\\
&=& -\frac{D}{2} \int \frac{d k_+}{i (2 \pi)^2}\,
     k_+\,(k_+-p_+)\,
     \int dk_- \frac{1}{k_- + i \epsilon/k_+}\,
     \frac{1}{k_- - p_- +i\epsilon/(k_+ -p_+)}. \label{eq:ex}
\end{eqnarray}
It is possible to carry out the integration over $k_-$ first. If we
take $p_+ > 0$, the poles in $k_-$ space are on opposite sides of the real
axis only in the case $0< k_+ < p_+$, then the $k_-$ integral is given
by the residue. It vanishes when $k_+ <0$ or $k_+>p_+$. Therefore
\begin{eqnarray}
&& \hspace{-11em}= -\frac{D}{4 \pi p_-} \int_0^{p_+} dk_+
k_+(k_+ - p_+)
\nonumber\\
&& \hspace{-11em}=
\frac{D}{24 \pi}\,\frac{p_+^3}{p_-}.
\end{eqnarray}

Evaluating other contributions, we obtain the two-point function of
the energy momentum tensor up to one-loop order:
\begin{eqnarray}
\VEV{{\rm T~} T_{++} (p)\,T_{++}(-p)}
= \frac{1}{6 \pi}\left\{D-2-12\,\xi\,(\,\xi+1\,)\right\}\,\frac{p_+^3}{p_-}.
\label{eq:twoene+}
\end{eqnarray}
\noindent
In a similar way we can obtain
\begin{eqnarray}
\VEV{{\rm T~} T_{--}(p)\,T_{--}(-p)}
= \frac{1}{6
\pi}\left\{D-2-12\,\xi\,(\,\xi+1\,)\right\}\,\frac{p_-^3}{p_+}.
\label{eq:twoene-}
\end{eqnarray}

This seems to suggest that the conformal symmetry is broken,
because the energy momentum tensor
may be coupled to the metric tensor.
It may be said that an arbitrary critical dimension $D$
may be allowed, simply by adjusting $\xi$.

However, this is not the case. We should consider the anomaly
more carefully. Since the metric field $h_{\alpha \beta}$ itself propagates
in the harmonic gauge, the two-point function of
the energy momentum tensor as defined in Eq.~(\ref{eq:eqmotion}) does
not correspond to the conformal anomaly directly. As seen later, we will
conclude that the anomaly disappears at $D=26$, as expected from
calculation in other gauges, by using an effective action.

\section{Conformal and Ghost Number Anomalies}
\reseteqnum

\noindent
We next consider an effective action. It is defined as
the Legendre transform of the connected vacuum functional $W$:
\begin{eqnarray}
&&\Gamma[\Phi;K] \equiv W[J;K] - \int d^2x J \Phi,
\nonumber\\
&&\Phi(x) \equiv \frac{\delta W}{\delta J(x)},
\hspace{2em}J \Phi \equiv J_{\alpha \beta} h^{\alpha \beta}
 + \overline{J}_\alpha c^\alpha + J^\alpha \overline{c}_\alpha
 + J_b^\alpha \left\{
 \widetilde{b}_\alpha+i\,(1-\xi)\, c^\beta \partial_\beta
  \overline{c}_\alpha
  \right\},
\nonumber
\end{eqnarray}
and
\begin{eqnarray}
{\rm exp}\,iW \equiv \int {\cal D}\Phi\,
{\rm exp}\,i \int d^2x \left[{\cal L} +J \Phi
  + K_{\alpha \beta}\, \delta_B h^{\alpha \beta}
  + K_\alpha\, \delta_B c^\alpha \right]. \label{eq:defeff}
\end{eqnarray}
We can easily find the Ward-Takahashi(W-T) identity
from the BRS invariance\,\cite{rf:DelMed}:
\begin{eqnarray}
\frac{\delta \Gamma}{\delta h^{\alpha \beta}}
\frac{\delta \Gamma}{\delta K_{\alpha \beta}}
+ \frac{\delta \Gamma}{\delta c^\alpha}
\frac{\delta \Gamma}{\delta K_\alpha}
+ \frac{\delta \Gamma}{\delta \overline{c}_\alpha} b_\alpha
=0. \label{eq:wt}
\end{eqnarray}
It should be noticed that the source term in Eq.~(\ref{eq:defeff})
contains not only $\widetilde{b}_\alpha$ but also a composite
operator of $i\,(1-\xi)c^\beta\,\partial_\beta\,\overline{c}_\alpha$.
This term is
indispensable for writing  down
the W-T identity for the BRS symmetry. Moreover it will play an important
role in calculating the anomaly.

The conformal mode $\phi$ does not propagate and has no interaction
with any fields at tree level. But we represent the W-T identity
including the conformal mode:
\begin{eqnarray}
\frac{\delta \Gamma_\phi}{\delta h^{\alpha \beta}}
\frac{\delta \Gamma_\phi}{\delta K_{\alpha \beta}}
+\frac{\delta \Gamma_\phi}{\delta \phi}
\frac{\delta \Gamma_\phi}{\delta K_\phi}
+ \frac{\delta \Gamma_\phi}{\delta c^\alpha}
\frac{\delta \Gamma_\phi}{\delta K_\alpha}
+ \frac{\delta \Gamma_\phi}{\delta \overline{c}_\alpha} b_\alpha
=0. \label{eq:wt2}
\end{eqnarray}
The definition of the effective action with the conformal mode
is the following:
\begin{eqnarray}
&&\hspace{-3em}
\Gamma_\phi[\Phi,\phi;K] \equiv W_\phi[J,J_\phi,K]
 -\int d^2x\left(J \Phi + J_\phi \phi \right), \nonumber\\
&&\hspace{-3em}
{\rm exp}\,iW_\phi \equiv \int {\cal D}\Phi\, {\rm
exp}\,i\int d^2x
\left[ {\cal L} + J\Phi+J_\phi\phi+K_{\alpha \beta}\,\delta_B h^{\alpha \beta}
+ K_\phi\,\delta_B \phi + K_\alpha\,\delta_B c^\alpha \right].
\end{eqnarray}

\subsection{Conformal Anomaly}
\begin{figure}[t]
\unitlength 2.8em
\begin{picture}(13,1.6)
\put(3.6,0.85){$\widetilde{b}_+(p)$}
\put(9.5,0.85){$\widetilde{b}_-(-p)$}
\put(5.4,1.3){$h_{--}$}
\put(8,1.3){$h_{++}$}
\centerline{
\epsfxsize=13em
\epsfbox{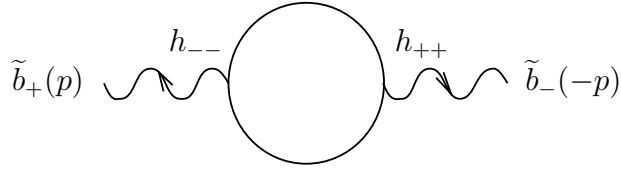}
}
\end{picture}
\caption{A graph contributing to a divergent integral.}
\label{fig:divb}
\end{figure}

\begin{figure}[t]
\unitlength 2.8em
\begin{picture}(13,2.5)(0,-0.5)
\put(4.3,-0.5){(a)}
\put(10.3,-0.5){(b)}
\put(1.6,0.95){$\widetilde{b}_+$}
\put(6,0.95){$i\,c_-\,\partial_+\overline{c}_+$}
\put(8.25,0.95){$i\,c_-\,\partial_+\overline{c}_+$}
\put(12,0.95){$i\,c_-\,\partial_+\overline{c}_+$}
\centerline{
\epsfxsize=27em
\epsfbox{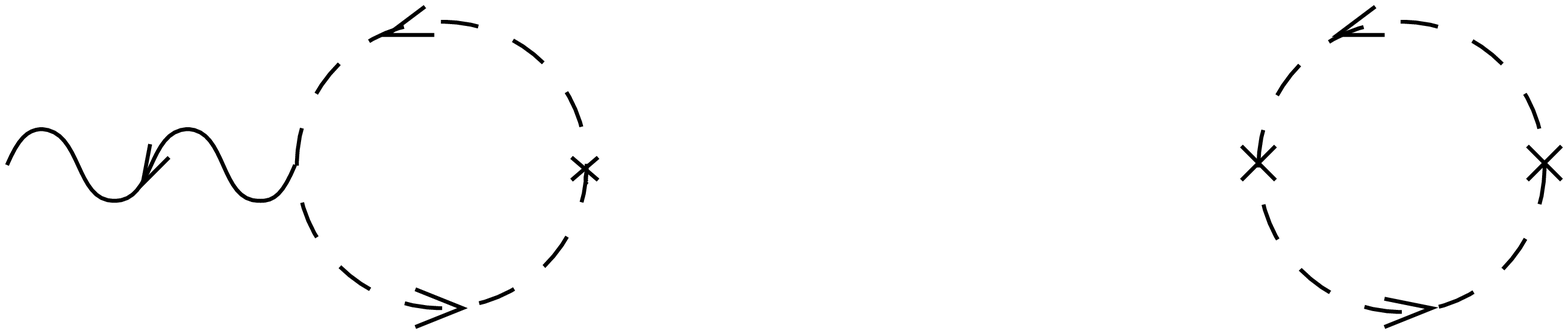}
}
\end{picture}
\caption{Diagrams contributing to the anomalous term in the effective
action. Figures (a) and (b) correspond to Eqs.~(3.8) and (3.9), respectively.}
\label{fig:bc}
\end{figure}

\noindent
Let us compute the two-point function of $b_\alpha$.
Through the same procedure as in the case of the energy momentum
tensor, it is given by
\begin{eqnarray}
\VEV{{\rm T~} \widetilde{b}_+ (p)\,\widetilde{b}_+(-p)}
= \frac{1}{96 \pi}\,\left\{D-2-12\,\xi\,(\,\xi+1\,)\right\}\,
\frac{p_+}{p_-}, \nonumber\\
\VEV{{\rm T~} \widetilde{b}_- (p)\, \widetilde{b}_-(-p)}
= \frac{1}{96 \pi}\,\,\left\{D-2-12\,\xi\,(\,\xi+1\,)\right\}\,
\frac{p_-}{p_+}. \label{eq:btild}
\end{eqnarray}
The results are finite. However the $+-$ component of the two-point function
is divergent. For instance, evaluating the graph in
Fig.~\ref{fig:divb},
we are faced with the integral as follows,
\begin{eqnarray}
-\frac{D}{8 p_+ p_-} \int \frac{dk_+ dk_-}{i (2 \pi)^2}\,
   \frac{k_+ k_-}{k_+ k_- +i \epsilon}\,
   \frac{(k_+ -p_+)(k_- -p_-)}{(k_+ -p_+)(k_- -p_-)+i \epsilon}.
\end{eqnarray}
Indeed the above integral is quadratically divergent and so it is
ill-defined.
Then it is necessary to
regularize it appropriately. We assume that a proper regularization
scheme exists and it is possible to evaluate a divergent integral by means
of it. We will take
the form of the $+-$ component as,
\begin{eqnarray}
\VEV{{\rm T~} \widetilde{b}_+(p)\,\widetilde{b}_-(-p)}
\equiv \frac{\widetilde{F}(p^2)}{p_+ p_-},    \label{eq:btild2}
\end{eqnarray}
where $\widetilde{F}(p^2)$ is a regular function of
$p^2$ under a certain
regularization scheme. This assumption
seems to be plausible for this integral.
Indeed $\widetilde{F}(p^2)$ becomes a local function, if we adopt the
regularization scheme in which the integral
is defined by using a momentum cut-off in Euclidean space.
But we have no need to specify the
regularization scheme further, because the term of
$\widetilde{F}(p^2)$ is irrelevant to the anomaly, as seen later.
We can compute the following two-point function in a
similar way,
\begin{eqnarray}
&&
\VEV{{\rm T~}\widetilde{b}_+(p)\,\, i\,c^\alpha \partial_\alpha
\overline{c}_+(-p)} = -\frac{1+2\,\xi}{16 \pi}\,\frac{p_+}{p_-},
\label{eq:twobc}\\
&&\VEV{{\rm T~}i\,c^\alpha \partial_\alpha
\overline{c}_+(p)\,\, i\,c^\beta \partial_\beta \overline{c}_+(-p)}
= -\frac{1}{8 \pi}\,\frac{p_+}{p_-}. \label{eq:twocc}
\end{eqnarray}
The Feynman graphs corresponding to Eqs.~(\ref{eq:twocc}) and
(\ref{eq:twobc}) are represented
in Fig.~\ref{fig:bc}.
 From Eqs.~(\ref{eq:btild}), (\ref{eq:twocc}) and (\ref{eq:twobc}),
we can obtain
\begin{eqnarray}
\VEV{{\rm T~} b_+(p)\,b_+(-p)}
&=& \frac{1}{96 \pi}\,\left\{D-2-12\,\xi\,(\,\xi+1\,)\right\}\,
\frac{p_+}{p_-}
-(1-\xi\,) \frac{1+2\,\xi}{16 \pi}\,\frac{p_+}{p_-}
\nonumber\\
&&-(1-\xi\,) \frac{1+2\,\xi}{16 \pi}\,\frac{p_+}{p_-}
-(1-\xi\,)^2 \frac{1}{8 \pi}\,\frac{p_+}{p_-}
\nonumber\\
&=& \frac{D-26}{96 \pi}\,\frac{p_+}{p_-}, \label{eq:twob} \\
\VEV{{\rm T~}b_-(p)\,b_-(-p)} &=& \frac{D-26}{96 \pi}\,\frac{p_-}{p_+}.
\label{eq:twob2}
\end{eqnarray}
We obtain also
\begin{eqnarray}
\hspace*{-17em}
{}~\VEV{{\rm T~}b_+(p)\,b_-(-p)} = \frac{F(p^2)}{p_+p_-}, \label{eq:twob3}
\end{eqnarray}
where $F(p^2)$ is a local function under a proper regularization scheme.

These
results are exact ones, though, so far, we have been computing them at
one-loop order. Actually there is no other contribution to
these two-point functions, because the metric and auxiliary fields
propagate only to each other and not to themselves. Therefore the
anomaly derived from these results is exact. Namely, it does
not have any correction of higher order.

Hence we have been able to compute the two-point function of
$b_\alpha$. The composite term $i c^\beta \partial_\beta
\overline{c}_\alpha$
is essential for the derivation of Eqs.~(\ref{eq:twob}) and
(\ref{eq:twob2}). The results of Eqs.~(\ref{eq:twob}) and
(\ref{eq:twob2}) infer that the BRS symmetry
is exact
if and only if $D=26$, and otherwise it is anomalous. Because
the left-hand side of Eqs.~(\ref{eq:twob}) and (\ref{eq:twob2}) must
be zero from the BRS
symmetry, $b_\alpha= -i\,\delta_B \overline{c}_\alpha$.

This fact can be shown more explicitly in the effective action. First,
we find
\begin{eqnarray}
\VEV{{\rm T~} b_+(p)\,h_{--}(-p)}=-\frac{i}{p_+}, \hspace{2em}
\VEV{{\rm T~} b_-(p)\,h_{++}(-p)}=-\frac{i}{p_-}.
\label{eq:bh}
\end{eqnarray}
The inverse of the two-point function of Eqs.~(\ref{eq:twob}),
(\ref{eq:twob2}), (\ref{eq:twob3}) and (\ref{eq:bh}) gives the
second order part of the fields
in the effective action:
\jot=8pt
\begin{eqnarray}
\Gamma[\Phi;K] = \int\,\,\hspace{-2em}&& d^2p\,\left[~i p_-\,h_{++}(p)\,b_-(-p)
 + ip_+\,h_{--}(p)\,b_+(-p) \frac{}{}\right. \nonumber\\
&&~\,+ \frac{D-26}{192 \pi}\left\{h_{++}(p)\,h_{++}(-p)\,\frac{p_-^3}{p_+}
+ h_{--}(p)\,h_{--}(-p)\,\frac{p_+^3}{p_-}\right\}\nonumber\\
&&\left.\frac{}{}+F(p^2)\,h_{++}(p)\,h_{--}(-p) + \cdots \right].
\label{eq:efac1}
\end{eqnarray}
\jot=6pt
Notice that $D-26$ of the two-point function of $b_\alpha$ gives
the factor for the nonlocal term of the metric field in the effective
action. As stated before, the $F(p^2)$ term can be removed and hence
not contribute to the anomaly, by using the freedom to add arbitrary local
counterterms to the effective action.

We will try to obtain a gauge invariant effective action, which
satisfies the W-T identity of Eq.~(\ref{eq:wt}). From
Eqs.~(\ref{eq:brstrans}) and
(\ref{eq:expmet}), the BRS transformation for
$h_{\alpha \beta}$ and $\phi$ can be derived:\vspace{-4ex}\\
\begin{eqnarray}
&&\delta_B h^{\alpha \beta} = - \partial^{(\alpha} c^{\beta)}
 + h^{(\alpha \gamma} \partial_\gamma c^{\beta)}
 - \partial_\gamma h^{\alpha \beta} c^\gamma
 + (\eta^{\alpha \beta} - h^{\alpha \beta})
  (\partial_\gamma c^\gamma - h^{\gamma \delta} \partial_\gamma
 c_\delta),  \nonumber\\
&&\delta_B \phi = \partial_\alpha c^\alpha - h^{\alpha \beta}
 \partial_\alpha c_\beta - \partial_\alpha \phi\,c^\alpha .
\end{eqnarray}
We rewrite this as follows,\vspace{-4ex}\\
\begin{eqnarray}
&&\delta_B h_{++} = -2\, \partial_+ c_+ ,\hspace{15em}\nonumber\\
&&\delta_B h_{--} = -2\, \partial_- c_- ,\nonumber\\
&&\delta_B \phi = \partial_+ c_- + \partial_- c_+,  \label{eq:brslin}
\end{eqnarray}
where terms higher than second order are omitted. Hereafter we
adopt this approximation. Considering the
dimension and ghost number of fields and sources, with the help of
Eq.~(\ref{eq:brslin}), we obtain
\begin{eqnarray}
\frac{\delta \Gamma}{\delta K_{--}(-p)} = 2i p_+ c_+(p),
\hspace{2em}
\frac{\delta \Gamma}{\delta K_{++}(-p)} = 2i p_- c_-(p).
\label{eq:wtlin1}
\end{eqnarray}
As easily seen by using Eq.~(\ref{eq:wtlin1}), it is
impossible to make the effective action satisfy the W-T identity of
Eq.~(\ref{eq:wt}),
if we add
only local counterterms of $h_{\alpha \beta}$ to it.
It is the nonlocal terms in the effective action which violate the
realization of the W-T identity and which are nonvanishing unless $D=26$.
Therefore the
nonlocal terms can be interpreted as the anomaly for the BRS
symmetry.

The BRS anomaly presented above is replaced with the conformal
anomaly, if we allow local terms with the conformal mode to be added
to the effective action and consider the W-T identity of
Eq.~(\ref{eq:wt2}).
The effective action $\Gamma_\phi$ satisfies, of course,
Eq.~(\ref{eq:wtlin1}), with $\Gamma$ replaced by $\Gamma_\phi$, and, in
addition, the following equation,
\begin{eqnarray}
\frac{\delta \Gamma_\phi}{\delta K_\phi(-p)}
= -i p_+ c_-(p) -i p_- c_+(p) \label{eq:wtlin2}.
\end{eqnarray}
Adding some local counterterms to Eq.~(\ref{eq:efac1}), we can form an
effective action
\jot=8pt
\begin{eqnarray}
\Gamma_\phi[\Phi, \phi;K] = \frac{D-26}{192 \pi}&&\hspace{-1.8em}\int d^2 p\,
\left[\,\,
\frac{p_-^3}{p_+}\,h_{++}(p)\,h_{++}(-p)
+\frac{p_+^3}{p_-}\,h_{--}(p)\,h_{--}(-p)\,
\right.
\nonumber\\
&&\hspace{-3em}+4\,p_-^2\, h_{++}(p)\, \phi(-p) + 4\,p_+^2\,
h_{--}(p)\,\phi(-p)
\nonumber\\
&&\hspace{-3.5em}\left.\frac{}{}+2\,p_+ p_-\,h_{++}(p)\,h_{--}(-p) +4\,
p_+p_-\,\phi(p)\,\phi(-p)\,\,\right]
+\cdots. \label{eq:efac2}
\end{eqnarray}
\jot=6pt
The nonanomalous terms and source terms have been omitted in this equation.
Indeed this effective action satisfies the W-T identity
of Eq.~(\ref{eq:wt2}).

Eq.~(\ref{eq:efac2}) is rewritten as
\begin{eqnarray}
\Gamma_\phi[\Phi,\phi;K] = \frac{D-26}{192 \pi}\int d^2p
\frac{1}{p_+p_-}\,
R(p)R(-p)+\cdots,  \label{eq:efac3}
\end{eqnarray}
where we have defined
\begin{eqnarray}
R(p) = -p_-^2\,h_{++}(p) - p_+^2\,h_{--}(p) -2\,p_+\,p_-\,\phi(p),
\end{eqnarray}
which is the expanded form of the scalar curvature by
Eq.~(\ref{eq:exp}).
Eq.~(\ref{eq:efac3}) corresponds to the conformal anomaly.
More explicitly, we can find it as
\begin{eqnarray}
\frac{\delta \Gamma_\phi}{\delta \phi(-p)} = \frac{D-26}{48 \pi}\,R(p).
\label{eq:conano}
\end{eqnarray}
At tree level, the dependence of the conformal mode is absent in the
action. It does not appear in the
effective action explicitly in the one loop calculation either. However, as is
a well-known fact\,\cite{rf:AlWitt}, if we require
the effective action to be invariant for the BRS symmetry, the
counterterms depending on the conformal mode are necessary for
recovering
the symmetry. As a result, we are led to the conformal anomaly as
given by Eq.~(\ref{eq:conano}) and the fact that the conformal
anomaly disappears only when $D=26$. Therefore the conclusion is that
the critical dimension is $26$ in the harmonic gauge.

\subsection{Ghost Number Anomaly}
\noindent
The Lagrangian is invariant under the following transformation,
\begin{eqnarray}
c_\alpha \longrightarrow {\rm e}^\theta\,c_\alpha,\hspace{2em}
\overline{c}_\alpha \longrightarrow {\rm e}^{-\theta}\,
\overline{c}_\alpha,
\end{eqnarray}
where $\theta$ is a real parameter. This is the ghost number symmetry.
The current of it is defined as
\begin{eqnarray}
J_C^\alpha \equiv i\,\widetilde{g}^{\alpha\beta}\,\left(
\overline{c}_\gamma \,\partial_\beta c^\gamma
- \partial_\beta \overline{c}_\gamma\, c^\gamma \right)
-i\,\xi\,\partial_\gamma\widetilde{g}^{\gamma
\beta}\,\overline{c}_\beta\,c^\alpha.
\end{eqnarray}

We will consider the $J_C^\alpha$-inserted vertex function which is
defined as follows,
\begin{eqnarray}
{\rm exp}\,i \Gamma_C[\Phi;L] \equiv
\int {\cal D}c {\cal D}\overline{c}~{\rm exp}\,i\int d^2x\,
\left[\,{\cal L} + L_\alpha J_C^\alpha\, \right].
\end{eqnarray}
This vertex function is sufficient to find the ghost number
anomaly.
Suppose that we obtain the following result,
\begin{eqnarray}
\frac{\delta \Gamma_C}{\delta L_+(-p)} &=&
 -i \lambda\, \frac{p_-^2}{p_+}\, h_{++}(p), \nonumber\\
\frac{\delta \Gamma_C}{\delta L_-(-p)} &=&
 -i \lambda\, \frac{p_+^2}{p_-}\, h_{--}(p),   \label{eq:efgh}
\end{eqnarray}
where local terms are omitted and $\lambda$ denotes a certain constant. These
terms, if any, cannot be taken away from the vertex function.  By
multiplying the momenta, Eq.~(\ref{eq:efgh})
becomes
\begin{eqnarray}
-ip_+\, \frac{\delta \Gamma_C}{\delta L_+(-p)}
-ip_-\, \frac{\delta \Gamma_C}{\delta L_-(-p)}
= \lambda \left(- p_-^2\, h_{++}(p)-p_+^2\, h_{--}(p)\right). \label{eq:efgh2}
\end{eqnarray}
Adding local functions to the vertex function, we obtain the
following equation from Eq.~(\ref{eq:efgh2}),
\begin{eqnarray}
-i p_\alpha\, \frac{\delta \Gamma_C}{\delta L_\alpha(-p)}
= -i p_\alpha\, \VEV{{\rm T~} J_C^\alpha(p)}
= \lambda R(p).  \label{eq:ghano}
\end{eqnarray}
This equation means that the ghost number current may not be conserved
at the quantum level. Namely, the ghost number anomaly exists
unless $\lambda$ equals zero.

However, $\lambda$ equals zero, and there is no anomaly in the ghost number
current. Indeed, $\lambda$ is given by evaluating a one-loop diagram
of ghost fields:
\begin{eqnarray}
-i\lambda\,\frac{p_-^2}{p_+} &=& -\frac{i}{2}\,\int \frac{dk_+
dk_-}{i(2\pi)^2}\,
(2k_--p_-)\,\frac{k_-}{k_+k_- +i\epsilon}\,
\frac{k_--p_-}{(k_+-p_+)(k_- -p_-)+i\epsilon}\nonumber\\
&=& \frac{-i}{4\pi p_+}\,\int^{p_-}_0 dk_-\, (2k_- -p_-)\nonumber\\
&=&0.
\end{eqnarray}

We can naturally understand the absence of the ghost number
anomaly as follows:
 The integration of Eq.~(\ref{eq:ghano}) gives us
\begin{eqnarray}
\int_M d^2x ~\partial_\alpha \VEV{{\rm T~} J_C^\alpha}
&=& \lambda \int_M d^2x \sqrt{-g}R \nonumber\\
&=& 4\pi \lambda~\chi(M),  \label{eq:ghano2}
\end{eqnarray}
where $M$ denotes some two-dimensional manifold and $\chi(M)$ denotes
the Euler number
of $M$.
On the other hand, the left hand side of Eq.~(\ref{eq:ghano2}) must be
given by the difference between the zero mode number of the ghost
field and
that of the anti-ghost field. Now the theory has invariance under
the FP conjugation of Eq.~(\ref{eq:fpconj}), and so the zero mode numbers
of ghost and anti-ghost fields are equal to each other. Hence
$\lambda$ equals zero and the
ghost number anomaly does not exist in the harmonic gauge.

\section{Summary}
In this paper, we have derived the anomaly for the BRS and conformal
symmetry in the harmonic gauge by considering the effective action
instead of the energy
momentum tensor. These anomalies are independent of the gauge
parameter.
Also the critical dimension, which is 26, is well-defined,
as in other gauges. It is difficult to discuss the theory in the harmonic
gauge because it cannot be solved exactly, unlike in the conformal or the
light-cone gauge. Therefore we should treat it carefully, respecting
the BRS
symmetry, for example, using the W-T identity for the effective action.

We have shown also that the ghost number current has no anomaly in the
harmonic gauge. This
is a natural consequence due to FP conjugation.

\section*{Acknowledgements}
The author would like to thank T.~Kugo, S.~Yahikozawa and M.~Sato for
valuable discussions.
He also thanks A.~Bordner for careful reading of the manuscript.

\newpage
\noindent


\begin{thebibliography}{99}
\bibitem{rf:GGRT} P.~Goddard, J.~Goldstone, G.~Rebbi and C.B.~Thorn,
{\sl Nucl.~Phys.}\ {\bf B56} (1973) 109.
\bibitem{rf:KatOga} M.~Kato and K.~Ogawa, {\sl Nucl.~Phys.}\ {bf B212}
(1983) 443.
\bibitem{rf:Poly} A.M.~Polyakov, {\sl Phys.~Lett.}\ {\bf 103B} (1981) 207.
\bibitem{rf:BSZ} L.~Baulieu, W.~Siegel and B.~Zwiebach, {\sl
Nucl.~Phys.}\ {\bf B287} (1987) 93.
\bibitem{rf:SieZwi} W.~Siegel and B.~Zwiebach, {\sl Nucl.~Phys.}\ {\bf 288}
(1987) 332.
\bibitem{rf:Duse} D.W.~D$\ddot{\rm u}$sedau, {\sl Phys.~Lett.}\ {\bf
188B} (1987) 51.
\bibitem{rf:FRP} D.Z.~Freedman, J.I.~Latorre and K.~Pilch,
{\sl Nucl.~Phys.}\ {\bf B306} (1988) 77.
\bibitem{rf:KR} U.~Kraemmer and A.~Rebhan, {\sl Nucl.~Phys.}\ {\bf B315}
(1989) 717.
\bibitem{rf:AbeNaka} M.~Abe and N.~Nakanishi,
{\sl Mod.~Phys.~Lett.}\ {\bf
A7} (1992) 1799.
\bibitem{rf:Kawano} T.~Kawano, {\sl Prog.~Theor.~Phys.}\ {\bf 93} (1995) 455.
\bibitem{rf:fuji} K.~Fujikawa, {\sl Phys.~Rev.}\ {\bf D25} (1982) 2584.
\bibitem{rf:NakaOji} For a review, N.~Nakanishi and I.~Ojima, {\it Covariant
Operator Formalism of Gauge Theories and Quantum Gravity}\ (World
Scientific, 1990).
\bibitem{rf:Naka} N.~Nakanishi, {\sl Prog.~Theor.~Phys.}\ {\bf 59} (1978)
2157.
\bibitem{rf:AlWitt} L.~Alvarez-Gaum$\acute{\rm e}$ and E.~Witten,
{\sl Nucl.~Phys.}\ {\bf B234} (1983) 269.
\bibitem{rf:tHooft} G.~'tHooft, {\sl Nucl.~Phys.}\ {\bf B75} (1974) 461.
\bibitem{rf:DelMed} R.~Delbourgo and M.R.~Medrano, {\sl Nucl.~Phys.}\
{\bf B110} (1976) 467.
\end{thebibliography}
\end{document}